\documentclass[aps,prl,reprint,groupedaddress, preprintnumbers]{revtex4-1}

\usepackage{amsfonts}
\usepackage{amsmath}
\usepackage{graphicx}
\usepackage[usenames]{color}
\usepackage[colorlinks=true, urlcolor=navyblue, linkcolor=navyblue, citecolor=navyblue]{hyperref}
\usepackage{lineno}
\usepackage[normalem]{ulem}
\usepackage{titlesec}
\titlespacing*{\section}{0pt}{1.1\baselineskip}{\baselineskip}
\titlespacing*{\subsection}{0pt}{1.1\baselineskip}{\baselineskip}
\titlespacing*{\subsubsection}{0pt}{1.1\baselineskip}{\baselineskip}

\def\be{\begin{equation}}
\def\ee{\end{equation}}
\def\bea{\begin{eqnarray}}
\def\eea{\end{eqnarray}}

\definecolor{navyblue}{rgb}{0,0.08,0.45}

\definecolor{orange}{rgb}{0.89,0.34,0.0}

\definecolor{green}{rgb}{0,0.6,0.4}

\newcommand{\half}{{\frac{1}{2}}}
\newcommand{\req}[1]{(\ref{#1})}
\newcommand{\mbf}[1]{\mathbf{#1}}

\begin{document}

\preprint{JLAB-THY-18-2630}
\preprint{SLAC-PUB-17217}


\title{Universality of Generalized Parton Distributions in Light-Front Holographic QCD}

\author{Guy F. de T\'eramond$^{1}$, Tianbo Liu$^{2,3}$, Raza Sabbir Sufian$^{2}$, \\ 
Hans G\"{u}nter Dosch$^{4}$, Stanley J. Brodsky$^{5}$,   Alexandre Deur$^{2}$}
\affiliation{
$^{1}$\mbox{Universidad de Costa Rica, 11501 San Jos\'e, Costa Rica}
$^{2}$\mbox{Thomas Jefferson National Accelerator Facility, Newport News, VA 23606, USA}
$^{3}$\mbox{Department of Physics, Duke University, Durham, NC 27708, USA}
$^{4}$\mbox{Institut f\"{u}r Theoretische Physik der Universit\"{a}t,  D-69120 Heidelberg, Germany}
$^{5}$\mbox{SLAC National Accelerator Laboratory, Stanford University, Stanford, CA 94309, USA}
}

\collaboration{HLFHS Collaboration}

 

\begin{abstract}

The structure of generalized parton distributions is determined from light-front holographic QCD up to a universal reparametrization function $w(x)$ which incorporates Regge behavior at small $x$ and inclusive counting rules at $x \to 1$. A simple ansatz for $w(x)$ which fulfills these physics constraints with a single-parameter results in precise descriptions of both the nucleon and the pion quark distribution functions in comparison with global fits. The analytic structure of the amplitudes leads to a connection with the Veneziano model and hence to a nontrivial connection with Regge theory and the hadron spectrum.

\end{abstract}



\maketitle

\section{INTRODUCTION}

Generalized parton distributions (GPDs)~\cite{Mueller:1998fv, Radyushkin:1996nd, Ji:1996ek} have emerged as a comprehensive tool to describe the nucleon structure as probed in hard scattering processes. GPDs link nucleon form factors (FFs) to longitudinal parton distributions (PDFs), and their first moment provide the angular momentum contribution of the nucleon constituents to its total spin through Ji's sum rule~\cite{Ji:1996ek}. The GPDs also encode information of the three-dimensional spatial structure of the hadrons: The Fourier transform of the GPDs gives the transverse spatial distribution of partons in correlation with their longitudinal momentum fraction $x$~\cite{Burkardt:2000za}.

Since a precise knowledge of PDFs is required for the analysis and interpretation of the scattering experiments in the LHC era, considerable efforts have been made to determine PDFs and their uncertainties by global fitting collaborations such as MMHT~\cite{Harland-Lang:2014zoa}, CT~\cite{Dulat:2015mca}, NNPDF~\cite{Ball:2017nwa}, and HERAPDF~\cite{Alekhin:2017kpj}. Lattice QCD calculations are using different methods, such as path-integral formulation of the deep-inelastic scattering hadronic tensor~\cite{Liu:1993cv,Liu:1999ak,Liang:2017mye}, inversion method~\cite{Horsley:2012pz, Chambers:2017dov}, {\it quasi}-PDFs~\cite{Ji:2013dva,Lin:2014zya,Alexandrou:2015rja,Alexandrou:2016jqi,Chen:2017lnm}, {\it pseudo}-PDFs~\cite{Radyushkin:2017cyf,Orginos:2017kos} and lattice cross-sections~\cite{Ma:2017pxb} to obtain the $x$-dependence of the PDFs. The current status and challenges for a meaningful comparison of lattice calculations with the global fits of PDFs can be found in~\cite{Lin:2017snn}.

There has been recent interest in the study of parton distributions using the framework of light-front holographic QCD (LFHQCD), an approach to hadron structure based on the holographic embedding of light-front dynamics in a higher dimensional gravity theory, with the constraints imposed by the underlying superconformal algebraic structure~\cite{Brodsky:2006uqa, deTeramond:2008ht, deTeramond:2013it, deTeramond:2014asa, Dosch:2015nwa, Brodsky:2014yha, Zou:2018eam}. This effective semiclassical approach to relativistic bound-state equations in QCD captures essential aspects of the confinement dynamics which are not apparent from the QCD Lagrangian, such as the emergence of a mass scale $\lambda = \kappa^2$, a unique form of the confinement potential, a zero mass state in the chiral limit: the pion, and universal Regge trajectories for mesons and baryons.

Various models of parton distributions based on LFHQCD~\cite{Abidin:2008sb, Vega:2010ns, Gutsche:2013zia, Chakrabarti:2013gra, Sharma:2014voa, Dehghani:2015jva, Chakrabarti:2015lba, Maji:2015vsa,  Chakrabarti:2017tek,  Mondal:2016xsm, Maji:2016yqo, Traini:2016jko, Traini:2016jru, Gutsche:2016gcd, Maji:2017ill, Rinaldi:2017roc, Bacchetta:2017vzh, Nikkhoo:2017won, Mondal:2017wbf, Kumar:2017dbf, Chouika:2017dhe, Muller:2017wms}  use as a starting point the analytic form of GPDs found in Ref.~\cite{Brodsky:2007hb}. This simple analytic form incorporates the correct high-energy counting rules of FFs~\cite{Brodsky:1973kr, Matveev:ra} and the GPD's $t$-momentum transfer dependence. One can also obtain effective light-front wave functions (LFWFs)~\cite{Brodsky:2014yha, Brodsky:2011xx} which are relevant for the computation of FFs and PDFs, including polarization dependent distributions~\cite{Maji:2017ill, Gutsche:2016gcd, Nikkhoo:2017won}. LFWFs are also used to study the skewness $\xi$-dependence of the GPDs~\cite{Traini:2016jko, Rinaldi:2017roc, Mondal:2017wbf, Chouika:2017dhe, Muller:2017wms}, and other parton distributions such as the Wigner distribution functions~\cite{Chakrabarti:2017tek, Gutsche:2016gcd}. The downside of the above phenomenological extensions of the holographic model is the large number of parameters required to describe simultaneously PDFs and FFs for each flavor.

Motivated by our recent analysis of the nucleon FFs in LFHQCD~\cite{Sufian:2016hwn}, we extend here our previous results for GPDs and LFWFs~\cite{Brodsky:2007hb, Brodsky:2011xx}. Shifting the FF poles to their physical location~\cite{Sufian:2016hwn} does not modify the exclusive counting rules but modifies the slope and intercept of the Regge trajectory, and hence the analytic structure of the GPDs which incorporates the Regge behavior. As a result, the $x$-dependence of PDFs and LFWFs is modified. Furthermore, the GPDs are defined in the present context up to a universal reparametrization function; therefore, imposing further physically motivated constraints is necessary.

\section{GPDs IN LFHQCD}

In LFHQCD the FF for arbitrary twist-$\tau$ is expressed in terms of Gamma functions~\cite{Brodsky:2007hb, Brodsky:2014yha}, an expression which can be recast in terms of the Euler Beta function $B(u,v)$ as~\cite{Zou:2018eam}
\be \label{FEuler}
F_\tau(t) = \frac{1}{N_\tau} B\left(\tau-1,  \frac{1}{2}- \frac{t}{4 \lambda} \right),
\ee
where
\be \label{intB}
B(u,v)= \int_0^1  dy\, y^{u -1} \, (1-y)^{v -1},
\ee
and $B(u,v) =  B(v, u)=\frac{\Gamma(u) \Gamma(v)}{\Gamma(u + v)}$ with $N_\tau = \sqrt{\pi} \frac{\Gamma(\tau -1)}{\Gamma\left(\tau - \half\right)}$. For fixed $u$ and large $v$ we have $B(u,v) \sim \Gamma(u) v^{-u}$: We thus
recover, for large $Q^2 = - t$, the hard-scattering scaling behavior~\cite{Brodsky:1973kr, Matveev:ra}
\be \label{FFas}
F_\tau(Q^2) \sim \left(\frac{1}{Q^2}\right)^{\tau-1} .
\ee

In contrast with the GPD twist which is determined by the quark-quark correlator,  twist-$\tau$
in \req{FEuler} and \req{FFas}  refers to the number of constituents in a given Fock component in the Fock expansion of the hadron state. It  controls the short distance behavior of the hadronic state and thus the power-law asymptotic behavior \req{FFas}.

For integer $\tau$ Eq.~\req{FEuler} generates the pole structure~\cite{Brodsky:2007hb}
\be\label{Ftau}
 F_\tau(Q^2) =  \frac{1}{{\Big(1 + \frac{Q^2}{M^2_0} \Big) }
 \Big(1 + \frac{Q^2}{M^2_{1}}  \Big)  \cdots
 \Big(1  + \frac{Q^2}{M^2_{{\tau-2}}} \Big)} ,
\ee
with $M_n^2 = 4  \lambda  \left(n+\half \right), \; n=0, 1, 2 \cdots {\tau-2}$, corresponding to the $\rho$ vector meson and its radial excitations~\cite{Brodsky:2014yha}. Notice that the Beta function in \req{FEuler} can be rewritten as $B\big(\tau - 1, 1 - \alpha(t)\big)$ with Regge trajectory
\be \label{VMRT}
\alpha(t) = \frac{t}{4 \lambda} + \half,
\ee
slope $\alpha' = \frac{1}{4 \lambda}$ and intercept $\alpha(0) = \half$.  This is just the $\rho$ trajectory emerging from LFHQCD.  The value of the universal scale $\lambda$ is fixed from the $\rho$ mass: $\sqrt{\lambda} = \kappa = m_\rho/ \sqrt{2} = 0.548$ GeV~\cite{Brodsky:2014yha, Brodsky:2016yod}.

Notice that the form factor \req{FEuler} can be expressed as a Veneziano amplitude~\cite{Veneziano:1968yb} $B\big(1 - \alpha(s), 1 - \alpha(t)\big)$, where the $s$-channel dependence is replaced by a fixed pole, $1 - \alpha(s) \to \tau -1$, allowed by unitarity constraints, since no resonances are formed in the $s$-channel~\cite{Ademollo:1969wd, Bender:1970ew, Landshoff:1970ce}

It will be useful to rewrite \req{FEuler} using the  reparametrization invariance  of the Euler Beta function~\req{intB}, and thus transform the integral representation of the form factor \req{FEuler} into the invariant form
\be
F_\tau(t) = \frac{1}{N_\tau}  \int_0^1 dx \, w'(x) w(x)^{- \frac{t}{4 \lambda} - \half} \big[1- w(x)\big]^{\tau-2},
\ee
if $w(x)$ is a monotonously  increasing function with fixed values at the integration limits  given by the constraints:
\be \label{wA}
 w(0) = 0, \quad w(1) = 1, \quad  w'(x) \ge 0,
\ee
 with $x \in [0,1]$. Any function $w(x)$ which satisfies the constraints \req{wA} will give the same result for the form factor.

Writing the flavor FF in terms of the valence GPD $F^q(t) = \int_0^1 dx\,  H^q_{\rm v}(x,t)$ at zero skewness, $ H^q(x,t) \equiv  H^q(x, \xi = 0, t)$, we obtain
\bea \label{GPDexp}
H^q(x,t) &=&\frac{1}{N_\tau}[1-w(x)]^{\tau-2}w(x)^{-\frac{1}{2}}w'(x) \, e^{\frac{t}{4\lambda}\log\big(\frac{1}{w(x)}\big)}\nonumber\\
&=&q_\tau(x)\exp[tf(x)],
\eea
where the PDF $q_\tau(x)$ and the profile function $f(x)$
\bea \label{qx} 
q_\tau(x) &=& \frac{1}{N_\tau} \big(1- w(x)\big)^{\tau-2}\, w(x)^{- \half}\, w'(x), \\
 \label{fx}
f(x) &=& \frac{1}{4 \lambda} \log\left(\frac{1}{w(x)}\right), 
\eea
are expressed in terms of the function $w(x)$ fulfilling conditions \req{wA}.

If for $x\sim0$, $w(x)$ behaves as $w(x) \sim x$, we find the $t$-dependence 
\be \label{wR0} 
   H^q_v(x,t) \sim x^{- t/ 4 \lambda } \, q_{\rm v}(x),
\ee
which is the Regge theory motivated ansatz for small-$x$ given in Ref.~\cite{Goeke:2001tz}  for $\alpha' =  \frac{1}{4 \lambda}.$

To study the behavior of $w(x)$ at large-$x$ we perform a Taylor expansion near $x =1$:
\be \label{wexp}
w(x) = 1 - (1-x) w'(1) + \half (1 - x)^2 w''(1) + \cdots .
\ee
Upon substitution of \req{wexp} in \req{qx} we find that the leading term in the expansion, which behaves as $(1-x)^{\tau-2}$, vanishes if $w'(1) = 0$. Hence setting 
\be \label{wp1}
w'(1) = 0 \quad {\rm and} \quad w''(1) \ne 0,
\ee
we find $q_\tau(x)  \sim (1-x)^{2 \tau - 3}$, which is precisely the  Drell-Yan inclusive counting rule at $x \to 1$~\cite{Drell:1969km, Blankenbecler:1974tm, Brodsky:1979qm}, corresponding to the form factor behavior~\label{FFasy} at large $Q^2$ \req{FFas}.

From Eq. \req{fx} it follows that the conditions \req{wp1} are equivalent to $f'(1) = 0$ and $f''(1) \ne 0$.
Since $\log(x) \sim 1-x$ for $x \sim 1$,  a simple ansatz for $f(x)$ consistent with  \req{wA}, \req{wR0} and \req{wp1} is
\be \label{fax} 
f(x) =   \frac{1}{4 \lambda}\left[  (1-x) \log\left(\frac{1}{x}\right) + a (1 - x)^2 \right],
\ee 
with $a$ being a flavor independent parameter. From \req{fx}
\be \label{wax}
w(x) = x^{1-x} e^{-a (1-x)^2},
\ee
an expression which incorporates Regge behavior at small-$x$  and inclusive counting rules  at large-$x$.

\subsection{Nucleon GPDs}

The nucleon GPDs are extracted from nucleon FF data~\cite{Diehl:2004cx, Guidal:2004nd, Selyugin:2009ic, Diehl:2013xca, GonzalezHernandez:2012jv} choosing specific $x$- and $t$-dependences of the GPDs for each flavor. One then finds the best fit reproducing the measured FFs and the valence PDFs. In our analysis of nucleon FFs~\cite{Sufian:2016hwn}, three free parameters are required: These are $r$, interpreted as an SU(6) breaking effect for the Dirac neutron FF, and $\gamma_p$ and $\gamma_n$, which account for the probabilities of higher Fock components (meson cloud), and are significant only for the Pauli FFs. The hadronic scale $\lambda$ is fixed by the $\rho$-Regge trajectory~\cite{Brodsky:2014yha}, whereas the Pauli FFs are normalized to the experimental values of the anomalous magnetic moments.

\subsubsection{Helicity Non-Flip Distributions}

Using the results from~\cite{Sufian:2016hwn} for the Dirac flavor FFs, we write the spin non-flip valence GPDs $H^q(x,t) = q(x) \exp \left[t f(x)\right]$ with
\bea
u_{\rm v}(x) &=& \left(2-\frac{r}{3}\right)q_{\tau=3}(x) +\frac{r}{3} \, q_{\tau=4}(x), \label{ux}\\
d_{\rm v}(x) &=&  \left(1-\frac{2 r}{3}\right) q_{\tau=3}(x) +\frac{2 r}{3} \, q_{\tau=4}(x) \label{dx} ,
\eea
for the $u$ and $d$ PDFs normalized to the valence content of the proton:
$ \int_0^1 dx \, u_{\rm v}(x) = 2$ and $\int_0^1 dx  \, d_{\rm v}(x)=1$. 
The PDF $q_\tau(x)$ and the profile function $f(x)$ are given by \req{qx} and \req{fx}, and $w(x)$ is given by \req{wax}. 
Positivity of the PDFs implies that $r \le 3/2$, which is smaller than the value $r=2.08$ found in~\cite{Sufian:2016hwn}. We shall use the maximum value $r = 3/2$, which does not change significantly our results in~\cite{Sufian:2016hwn}.

\begin{figure}[htbp] 
\begin{center} 
\includegraphics[width=8.2cm]{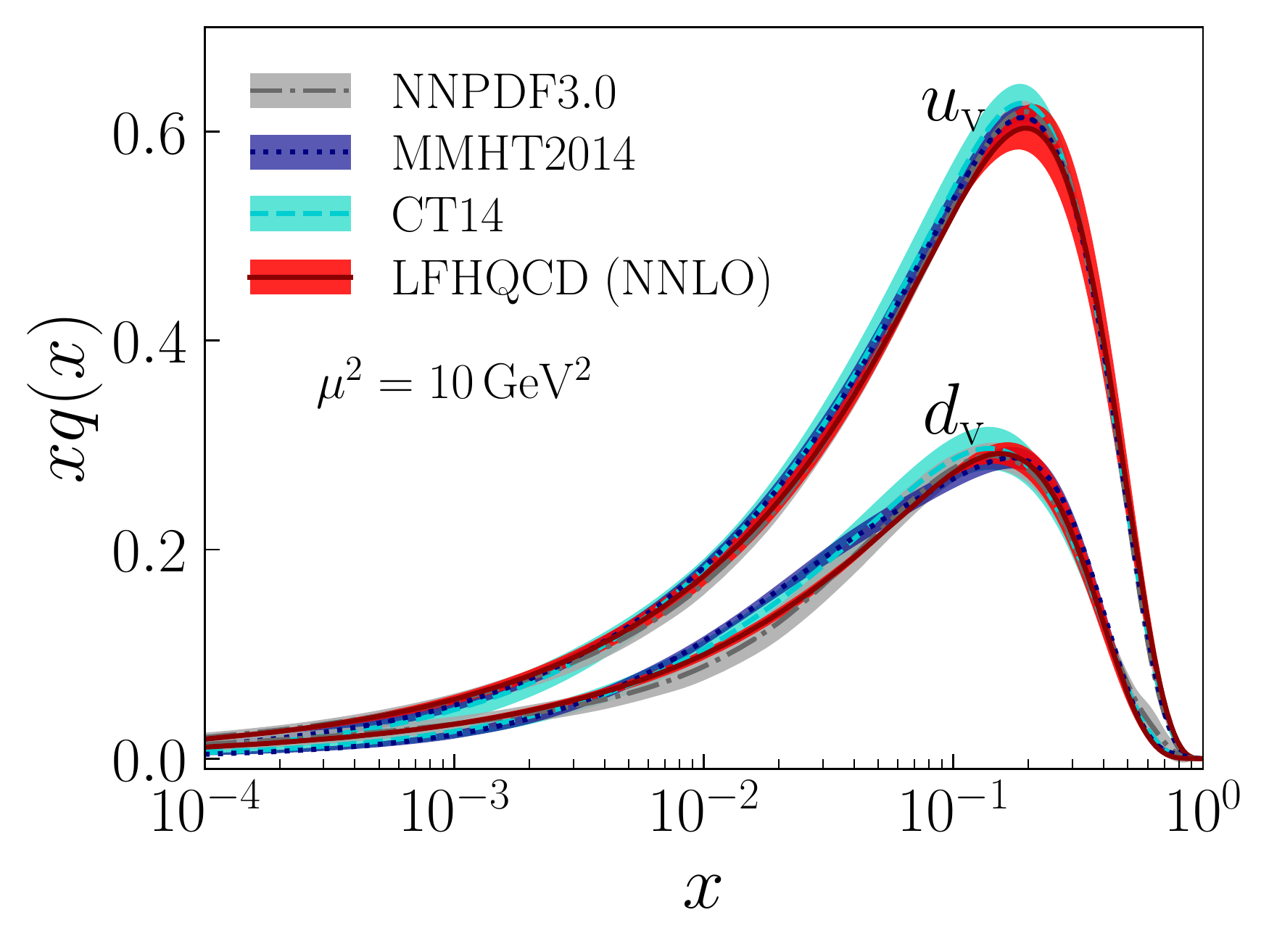}
\setlength\abovecaptionskip{-4pt}
\setlength\belowcaptionskip{-6pt}
\caption{\label{globalfit}  Comparison for $x q(x)$ in the proton from LFHQCD (red bands) and global fits: MMHT2014 (blue bands)~\cite{Harland-Lang:2014zoa}, CT14~\cite{Dulat:2015mca} (cyan bands), and NNPDF3.0 (grey bands)~\cite{Ball:2014uwa}. LFHQCD results are evolved from the initial  scale $\mu_0 = 1.06 \pm 0.15 \,\rm GeV$.}
\label{globalfits}
\end{center}
\end{figure}

\begin{figure}[!h]
\centering
\includegraphics[width=8.9cm]{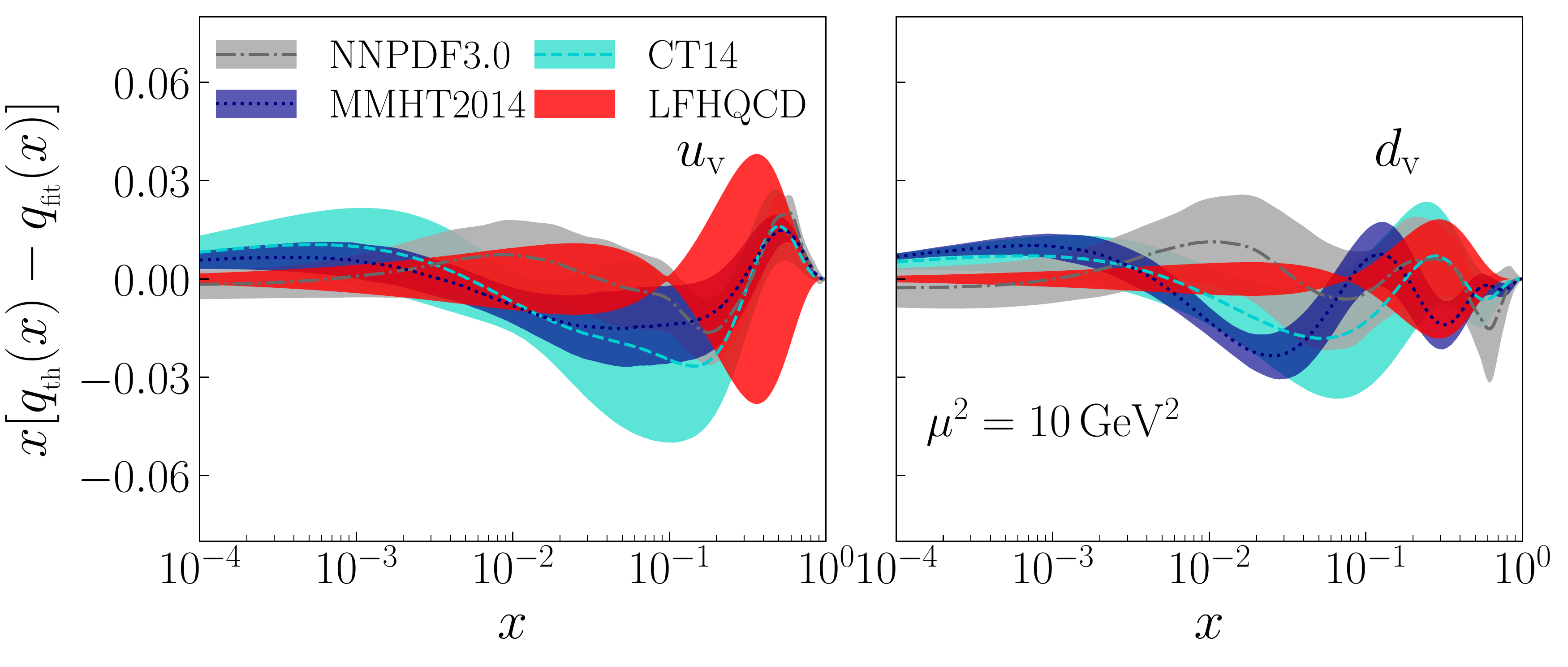}
\caption{\label{globalfitdiff}  Difference between our PDF results and global fits.}
\end{figure}

\begin{figure}[htbp] 
\begin{center} 
\includegraphics[width=8.8cm]{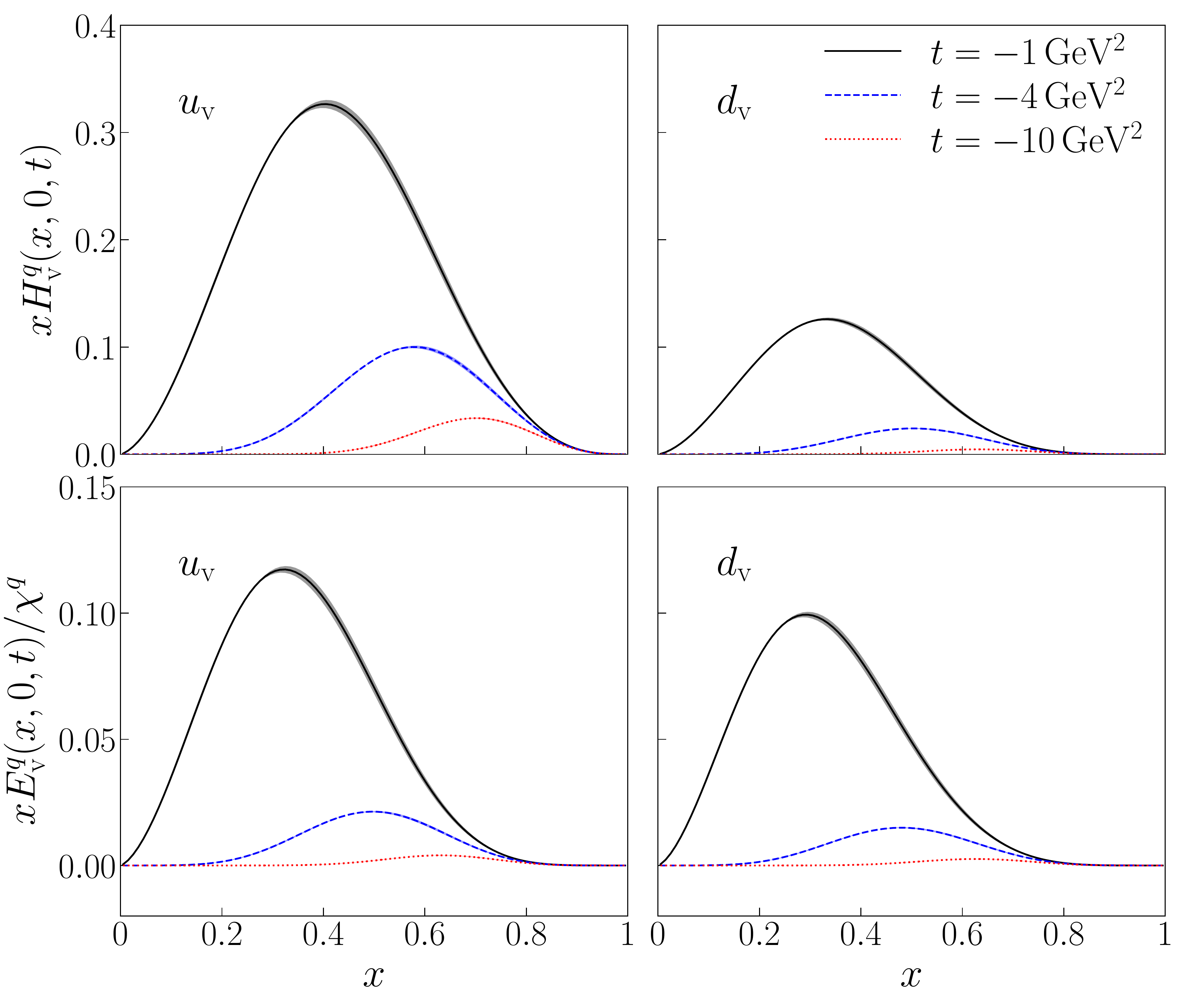}
\setlength\abovecaptionskip{-4pt}
\setlength\belowcaptionskip{-6pt}
\caption{\label{GPDs} Nucleon GPDs for different values of $- t =  Q^2$  at the scale $\mu_0=1.06\pm0.15\,\rm GeV$. Top: spin non-flip $H^q_{\rm v}(x,t)$. Bottom: spin-flip $E^q_{\rm v}(x,t)$.}
\end{center}
\end{figure}

The PDFs \req{ux} and \req{dx} are evolved to a higher scale $\mu$ with the Dokshitzer-Gribov-Lipatov-Altarelli-Parisi (DGLAP) equation~\cite{Altarelli:1977zs,Dokshitzer:1977sg,Gribov:1972ri} in the $\overline{\rm MS}$ scheme using the HOPPET toolkit~\cite{Salam:2008qg}. The initial scale is chosen at the matching scale between LFHQCD and pQCD as $\mu_0 = 1.06 \pm 0.15 \,\rm GeV$~\cite{Deur:2016opc} in the $\overline{\rm MS}$ scheme at next-to-next-to-leading order (NNLO). The strong coupling constant $\alpha_s$ at the scale of the $Z$-boson mass is set to $0.1182$~\cite{Patrignani:2016xqp}, and the heavy quark thresholds are set with $\overline{\rm MS}$ quark  masses as $m_c=1.28\,\rm GeV$ and $m_b=4.18\,\rm GeV$~\cite{Patrignani:2016xqp}. The PDFs are evolved to $\mu^2=10\,\rm GeV^2$ at NNLO to compare with the global fits by the  MMHT~\cite{Harland-Lang:2014zoa}, CT~\cite{Dulat:2015mca}, and NNPDF~\cite{Ball:2014uwa}  collaborations as shown in Fig.~\ref{globalfit}.  The value  $a = 0.531 \pm 0.037$ is determined from the first moment of the GPD,
$
\int_0^1 dx \,x  H^q_{\rm v}(x, t = 0) = A^q_{\rm v}(0)
$
from  the global data fits with average values $A_v^u(0) = 0.261 \pm 0.005$ and 
$A_v^d(0) = 0.109 \pm 0.005$. The model uncertainty (red band) includes the uncertainties in $a$ and $\mu_0$~\cite{mu0}.  We also indicate the  difference between our results and global fits in Fig.~\ref{globalfitdiff}. 
The $t$-dependence of $H^q_{\rm v}(x,t)$ is illustrated in Fig.~\ref{GPDs}.  Since our PDFs scale as $q(x) \sim x^{-1/2}$ for small-$x$, the Kuti-Weisskopf behavior for the non-singlet structure functions $F_{2p} (x) - F_{2n}(x) \sim x (u_v(x) - d_v(x)) \sim x^{1/2}$ is satisfied~\cite{Kuti:1971ph, Landshoff:1970ff}.

\subsubsection{Helicity-Flip Distributions}

The spin-flip GPDs $E^q_{\rm v}(x,t) = e^q_{\rm v}(x) \exp \left[t f(x)\right]$ follow from the flavor Pauli FFs in~\cite{Sufian:2016hwn} given in terms of twist-4 and twist-6 contributions
\be
e^q_{\rm v}(x) = \chi_q \left[\left(1 - \gamma_q \right) \,q_{\tau=4}(x) + \gamma_q \, q_{\tau=6}(x) \right],
\ee
normalized to the flavor anomalous magnetic moment $ \int_0^1 dx \, e^q_{\rm v}(x) = \chi_q$, with $\chi_u= 2 \chi_p + \chi_n = 1.673$ and $\chi_d =   2 \chi_n +\chi_p = -2.033$. The factors $\gamma_u$ and $\gamma_d$ are
\be
\gamma_u \equiv \frac{2 \chi_p \gamma_p + \chi_n \gamma_n}{2 \chi_p + \chi_n}, \quad 
\gamma_d \equiv \frac{2 \chi_n \gamma_n + \chi_p \gamma_p}{2 \chi_n + \chi_p}, 
\ee
where the higher Fock probabilities $\gamma_{p,n}$ represent the large distance pion contribution and have the values $\gamma_p=0.27$ and $\gamma_n=0.38$~\cite{Sufian:2016hwn}. Our results for $E^q_{\rm v}(x,t)$ are displayed in Fig.~\ref{GPDs}.

\begin{figure}[htbp] 
\begin{center} 
\includegraphics[width=8.2cm]{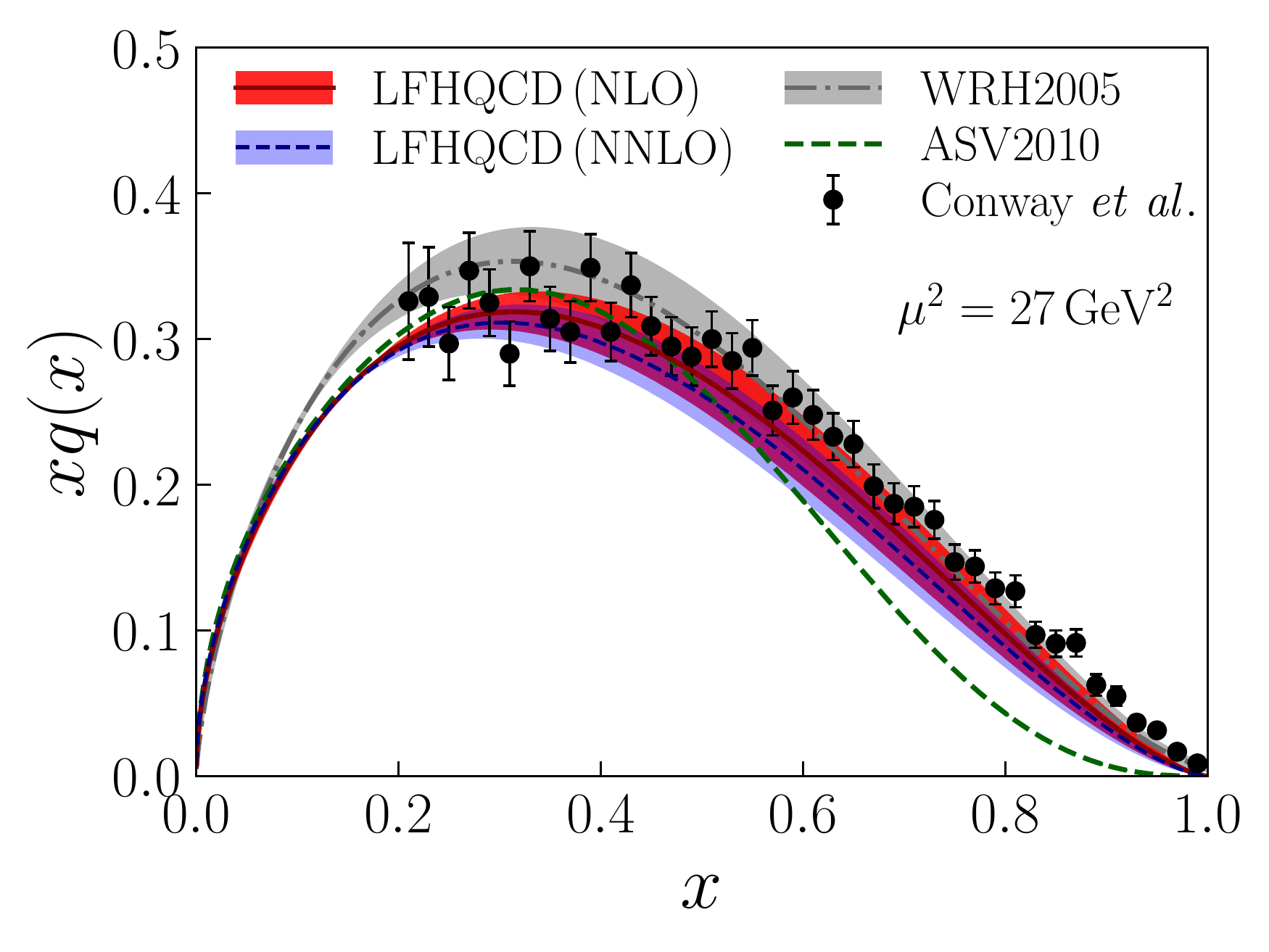}
\setlength\abovecaptionskip{-4pt}
\setlength\belowcaptionskip{-6pt}
\caption{\label{pionGPD} Comparison for $x q(x)$ in the pion from LFHQCD (red band) with the NLO fits~\cite{Wijesooriya:2005ir,Aicher:2010cb} (gray band and green curve) and the LO extraction~\cite{Conway:1989fs}. NNLO results are also included (light blue band). LFHQCD results are evolved from the initial scale $\mu_0 = 1.1 \pm 0.2 \,\rm GeV$ at NLO and the initial scale $\mu_0=1.06\pm0.15\,\rm GeV$ at NNLO.}
\end{center}
\end{figure}

\subsection{Pion GPD}

The expression for the pion GPD $H^{u, \bar d}_{\rm v}(x,t) = q_{\rm v}^{u, \bar d}(x) \exp \left[t f(x)\right]$ follows from the pion FF in~\cite{deTeramond:2010ez}, where the contribution from higher Fock components was determined from the analysis of the time-like region~\cite{deTeramond:2010ez}. Up to twist-4
\be
q^{u,\bar d}_{\rm v}(x) = (1 - \gamma) q_{\tau=2}(x) + \gamma q_{\tau=4}(x),\label{pionpdf}
\ee
where the PDFs are normalized to the valence quark content of the pion
$ 
\int_0^1 dx \, q^{u, \bar d}_{\rm v}(x) = 1
$,
and $\gamma = 0.125$ represents the meson cloud contribution  determined in~\cite{Brodsky:2014yha}.

\begin{figure}[htbp] 
\begin{center} 
\includegraphics[width=5.8cm]{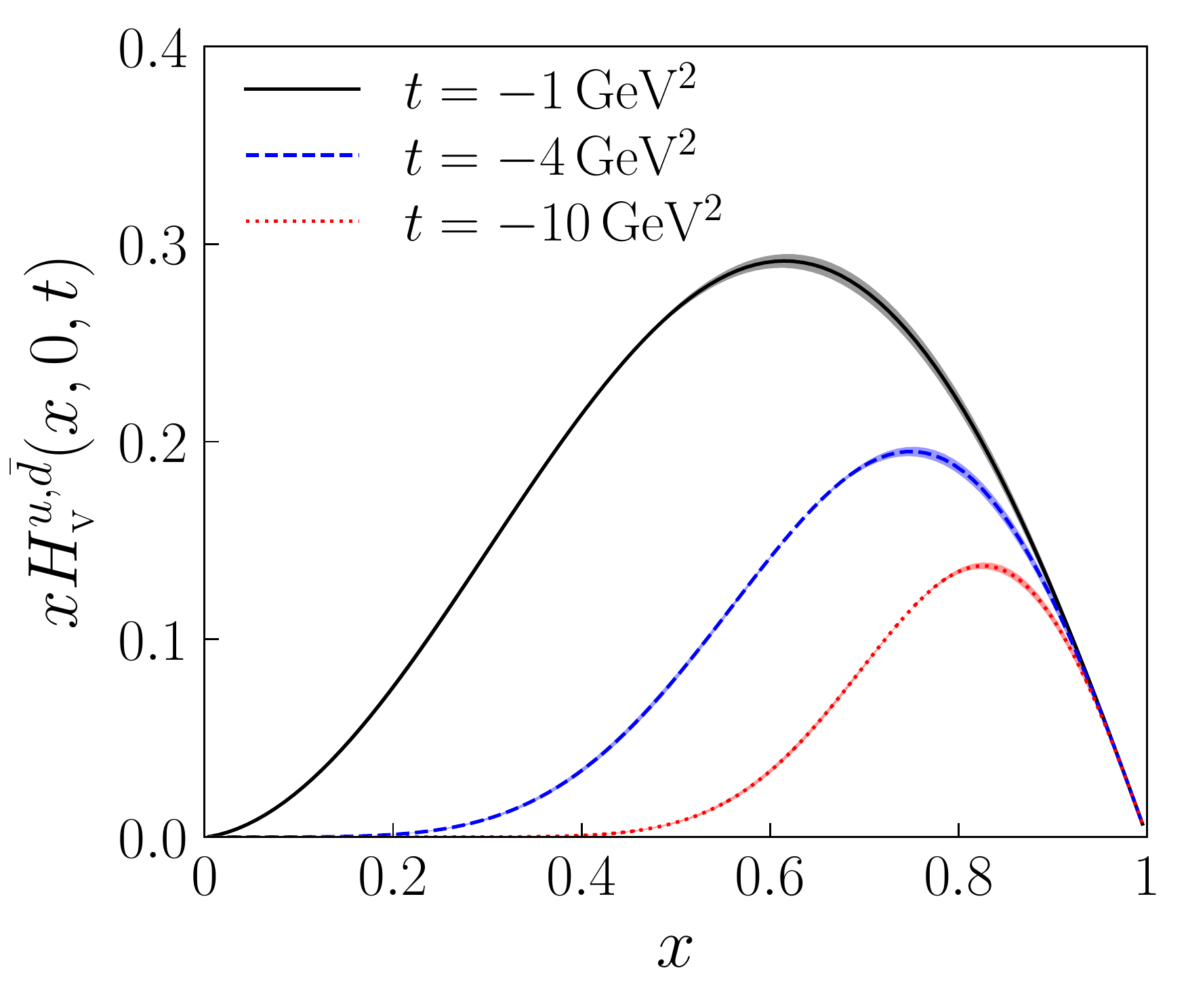}
\setlength\abovecaptionskip{-4pt}
\setlength\belowcaptionskip{-6pt}
\caption{\label{piGPDs}  Pion GPD for different values of $- t =  Q^2$  at the scale $\mu_0=1.1\pm0.2\,\rm GeV$.}
\end{center}
\end{figure}

The pion PDFs are evolved to $\mu^2=27\,\rm GeV^2$ at next-to-leading order (NLO) to compare with the NLO global analysis in~\cite{Wijesooriya:2005ir, Aicher:2010cb} of the data~\cite{Conway:1989fs}. The initial scale is set at $\mu_0 = 1.1 \pm 0.2 \,\rm GeV$ from the matching procedure in Ref.~\cite{Deur:2016opc} at NLO. The result is shown in Fig.~\ref{pionGPD}, and the $t$-dependence of  $H^{q}_{\rm v}(x,t)$ is illustrated in Fig.~\ref{piGPDs}. We have also included the NNLO results in Fig.~\ref{pionGPD}, to compare with future data analysis.

Our results are in good agreement with the data analysis in Ref.~\cite{Wijesooriya:2005ir} and consistent with the nucleon global fit results through the GPD universality described here. There is however a tension with the data analysis in~\cite{Aicher:2010cb} for $x \ge 0.6$ and with the Dyson-Schwinger results in~\cite{Chen:2016sno}  which incorporate the $(1-x)^2$ pQCD  falloff at large-$x$  from hard gluon transfer to the spectator quarks. In contrast, our nonperturbative results falloff as $1-x$ from the leading  twist-2 term in \req{pionpdf}. A softer falloff $\sim (1-x)^{1.5}$ in Fig.~\ref{pionGPD} follows from  DGLAP evolution.   Our analysis incorporates the nonperturbative behavior of effective LFWFs in the limit of zero quark masses. However, if we include a nonzero quark mass in the LFWFs~\cite{Brodsky:2014yha, Chabysheva:2012fe, Li:2017mlw}, the PDFs will be further suppressed at $x \to 1$.

\section{Effective LFWFs}

Form factors in light-front quantization can be written in terms of an effective single-particle density~\cite{Soper:1976jc}
\be
F(Q^2) = \int_0^1 dx \rho(x, Q),
\ee
where $\rho(x,Q) =2 \pi \int_0^\infty \!  db \,  b \, J_0\big(b Q (1-x)\big) \vert \psi_{\rm eff} (x,b)\vert^2$ with transverse separation $b = \vert \mbf{b}_\perp \vert$.
From~\req{GPDexp} we find the effective LFWF
\be \label{LFWFb}
\psi_{\rm eff}^\tau(x, \mbf{b}_\perp) = \frac{1}{2 \sqrt{\pi}} \sqrt{\frac{q_\tau(x)}{f(x)}} 
 (1-x) \exp \left[ - \frac{(1-x)^2 }{8 f(x) } \, \mbf{b}_\perp^2\right],
\ee
in the transverse impact space representation with $q_\tau(x)$ and $f(x)$ given by \req{qx} and \req{fx}. The normalization is
$\int_0^1 dx \int d^2\mbf{b}_\perp \left \vert  \psi_{\it eff}(x, \mbf{b}_\perp) \right\vert^2 = 1$, provided that $\int_0^1 dx \, q_\tau(x) = 1$.
In the transverse momentum space 
\be  \label{LFWFk}
\psi_{\rm eff}^\tau(x, \mbf{k}_\perp) 
= 8 \pi \frac{ \sqrt{q_\tau(x) f(x) }}{1-x} \,
\exp\left[ -  \frac{2 f(x)}{(1-x)^2} \, \mbf{k}_\perp^2 \right],
\ee
with normalization $\int_0^1 dx \int \frac{d^2\mbf{k}_\perp}{16 \pi^3} \left\vert  \psi_{\it eff}(x, \mbf{k}_\perp) \right\vert^2=1$.

\section{CONCLUSION AND OUTLOOK}

The results presented here for the GPDs provide a new nonperturbative structural framework for the exclusive-inclusive connection which is fully consistent with the LFHQCD results for the hadron spectrum. The PDFs are flavor-dependent and expressed as a superposition of PDFs $q_\tau(x)$ of different twist. In contrast, the GPD profile function $f(x)$ is universal. Both $q(x)$ and $f(x)$ can be expressed in terms of a universal reparametrization function $w(x)$, which incorporates Regge behavior at small-$x$ and  inclusive counting rules at large-$x$. A simple ansatz for $w(x)$, which satisfies all the physics constraints, leads to a precise description of parton distributions and form factors for the pion and nucleons in terms of a single physically constrained parameter.  In contrast with the eigenfunctions of the holographic LF Hamiltonian~\cite{Brodsky:2014yha}, the effective LFWFs obtained here incorporate the nonperturbative pole structure of the amplitudes, Regge behavior and exclusive and inclusive counting rules. The LFWFs  can be used to study other parton distributions, such as the transverse momentum dependent parton distributions and the Wigner distributions. The analytic structure of FFs and GPDs leads to a connection with the Veneziano amplitude  which incorporates the $\rho$ Regge trayectory determined in LFHQCD. It could give further insights in understanding the quark-hadron duality and hadron structure. The falloff of the pion PDF at large-$x$ is an unresolved issue~\cite{Holt:2010vj}.

\section{ACKNOWLEDGEMENTS}

HGD wants to thank Markus Diehl for valuable comments. GdT thanks Alessandro Bacchetta, Sabrina Cotogno and Barbara Pasquini for helpful remarks.  GdT and SJB thank Craig Roberts for helpful comments. This work is supported in part by the U.S. Department of Energy, Office of Science, Office of Nuclear Physics under contract No. DE-AC05-06OR23177 and No. DE-FG02-03ER41231, and by the Department of Energy, contract DE--AC02--76SF00515.

\end{document}